\documentclass[conference]{IEEEtran}
\usepackage{graphicx}
\usepackage{subcaption} 
\usepackage{tabularx}
\usepackage{hyperref}

\ifCLASSINFOpdf
\else
\fi
\begin{document}
%
\title{Empirical evaluation of LLMs in predicting fixes of Configuration bugs in Smart Home System}



%
\author{
    \IEEEauthorblockN{\textbf{Sheikh Moonwara Anjum Monisha (Author)}}
    \IEEEauthorblockA{
        Dept. Of Computer Science\\
        Blacksburg, VA\\
        msheikhmoonwaraa@vt.edu}
    \and
    \IEEEauthorblockN{\textbf{Atul Bharadwaj (Author)}}
    \IEEEauthorblockA{
        Dept. Of Computer Science\\
        Blacksburg, VA\\
        atulnm2002@vt.edu}
}



\maketitle

\begin{abstract}
This empirical study evaluates the effectiveness of Large Language Models (LLMs) in predicting fixes for configuration bugs in smart home systems. The research analyzes three prominent LLMs - GPT-4, GPT-4o (GPT-4 Turbo), and Claude 3.5 Sonnet - using four distinct prompt designs to assess their ability to identify appropriate fix strategies and generate correct solutions. The study utilized a dataset of 129 debugging issues from the Home Assistant Community, focusing on 21 randomly selected cases for in-depth analysis. Results demonstrate that GPT-4 and Claude 3.5 Sonnet achieved 80\% accuracy in strategy prediction when provided with both bug descriptions and original scripts. GPT-4 exhibited consistent performance across different prompt types, while GPT-4o showed advantages in speed and cost-effectiveness despite slightly lower accuracy. The findings reveal that prompt design significantly impacts model performance, with comprehensive prompts containing both description and original script yielding the best results. This research provides valuable insights for improving automated bug fixing in smart home system configurations and demonstrates the potential of LLMs in addressing configuration-related challenges.
\end{abstract}


%
\IEEEpeerreviewmaketitle

\section{Introduction}
The field of smart home automation has seen significant growth in recent years, with an increasing number of households adopting various Internet of Things (IoT) devices and systems to enhance their living environments. However, as these systems become more complex, users often encounter configuration bugs that can hinder the proper functioning of their smart home setups. This paper presents an empirical evaluation of the effectiveness of Large Language Models (LLMs) in predicting fixes for configuration bugs in smart home systems, focusing on the use of advanced AI models to address common issues faced by users\cite{Naveed2023ACO}.


\subsection{Smart Home Systems}

Smart home systems are networks of interconnected devices and appliances that can be remotely controlled and automated to enhance comfort, energy efficiency, and security in residential environments\cite{web3}. These systems typically include:
\begin{itemize}
    \item Smart lighting
    \item Thermostats and HVAC controls
    \item Security cameras and alarm systems
    \item Voice-activated assistants
    \item Smart appliances (e.g., refrigerators, washing machines)
    \item Automated window coverings and door locks
\end{itemize}
The global smart home market is projected to reach USD 338.28 billion by
2030, growing at a 20.1\% CAGR. Key drivers include increas-
ing internet use, rising disposable incomes, the need for remote
home monitoring, and demand for energy-efficient solutions.
Home Assistant (HA) stands out among various platforms for
its open-source, privacy-focused design, with over 332,000
active installations as of March 2024\cite{smartHomeMarket2032}. The widespread adoption
of HA presents a unique opportunity for software engineering
research, as (1) its data and programs offer a comprehensive
representation of smart home systems, and (2) our research
can contribute to the future development of smart home
technologies\cite{web3}\cite{web4}.

\subsection{Configuration Bugs}

Configuration bugs in smart home systems refer to errors or inconsistencies in the setup and programming of automation rules, device interactions, and system behaviors. These bugs can arise from various sources, including:
\begin{itemize}
    \item Incorrect syntax in automation scripts
    \item Mismatched trigger conditions
    \item Improper use of variables and templates
    \item Inconsistent state definitions
    \item Incompatible device integrations
\end{itemize}

Such bugs can lead to unintended system behaviors, reduced efficiency, or complete failure of automated processes, significantly impacting the user experience and the overall functionality of the smart home system.

\subsection{Large Language Models (LLMs)}

Large Language Models are advanced artificial intelligence systems trained on vast amounts of text data to understand and generate human-like language\cite{Naveed2023ACO}. These models have demonstrated remarkable capabilities in various natural language processing tasks, including:
\begin{itemize}
    \item Text generation
    \item Language translation
    \item Question answering
    \item Code generation and analysis
\end{itemize}

In the context of this study, we focus on three prominent LLMs:
\begin{itemize}
    \item GPT-4: Developed by OpenAI, GPT-4 is known for its advanced reasoning capabilities and versatility across a wide range of tasks\cite{achiam2023gpt}.
    \item GPT-4o (GPT-4 Turbo): An optimized version of GPT-4, designed for faster processing and improved cost-effectiveness in large-scale applications\cite{10.1145/3650105.3652291}\cite{hurst2024gpt}.
    \item Claude 3.5 Sonnet: Created by Anthropic, this model emphasizes safety and the ability to handle large contexts, making it suitable for sensitive tasks\cite{Enis2024FromLT}.
\end{itemize}

\subsection{Home Assistant Community (HAC)}

The Home Assistant Community\cite{web4} is a platform where users of the popular open-source home automation system, Home Assistant, share knowledge, discuss issues, and seek solutions for their smart home setups. It serves as a valuable resource for both novice and experienced users, providing a wealth of real-world examples of configuration challenges and their resolutions.

\subsection{Prompt Engineering}

Prompt engineering refers to the process of designing and optimizing input prompts for LLMs to elicit desired outputs. In this study, we explore four distinct prompt designs:
\begin{itemize}
    \item Description Only
    \item Description + Original Script
    \item Description + Original Script + Specific Issue
    \item Issue Only
\end{itemize}

These prompts are carefully crafted to provide varying levels of context and information to the LLMs, allowing us to assess their performance under different input conditions.

\subsection{Configuration Bug Resolution Strategies}

The study builds upon previous research that identified eight representative resolution strategies for common configuration bugs in smart home systems. These strategies encompass various aspects of automation configuration, including:
\begin{itemize}
    \item Proper use of quotes and string literals
    \item Correct indentation and formatting
    \item Appropriate trigger selection
    \item Accurate state reading and specification
    \item Efficient handling of multiple triggers and entity values
\end{itemize}

By mapping these strategies to the dataset and using them as a framework for evaluation, we can assess the LLMs' ability to identify and apply appropriate fixes to configuration issues.

\subsection{Significance of the Study}

This research aims to bridge the gap between advanced AI technologies and practical smart home automation challenges. By evaluating the capability of LLMs to predict and generate fixes for configuration bugs, we seek to:
\begin{itemize}
    \item Enhance the user experience for smart home enthusiasts by providing more accessible troubleshooting solutions
    \item Reduce the time and effort required to resolve common configuration issues
    \item Explore the potential for AI-assisted automation in smart home system maintenance and optimization
    \item Contribute to the broader field of AI applications in IoT and home automation
\end{itemize}

Through a rigorous empirical evaluation of different LLMs and prompt designs, this study offers valuable insights into the effectiveness of AI-driven approaches to solving real-world smart home configuration challenges. The findings presented here have implications for both end-users and developers in the smart home ecosystem, potentially paving the way for more intelligent and user-friendly automation systems in the future.

\section{Approach}
\subsection{Dataset and Resolution Strategies }

\subsubsection{Dataset Extraction} 
\begin{figure*}[h!]
    \centering
    \includegraphics[width=0.98\textwidth]{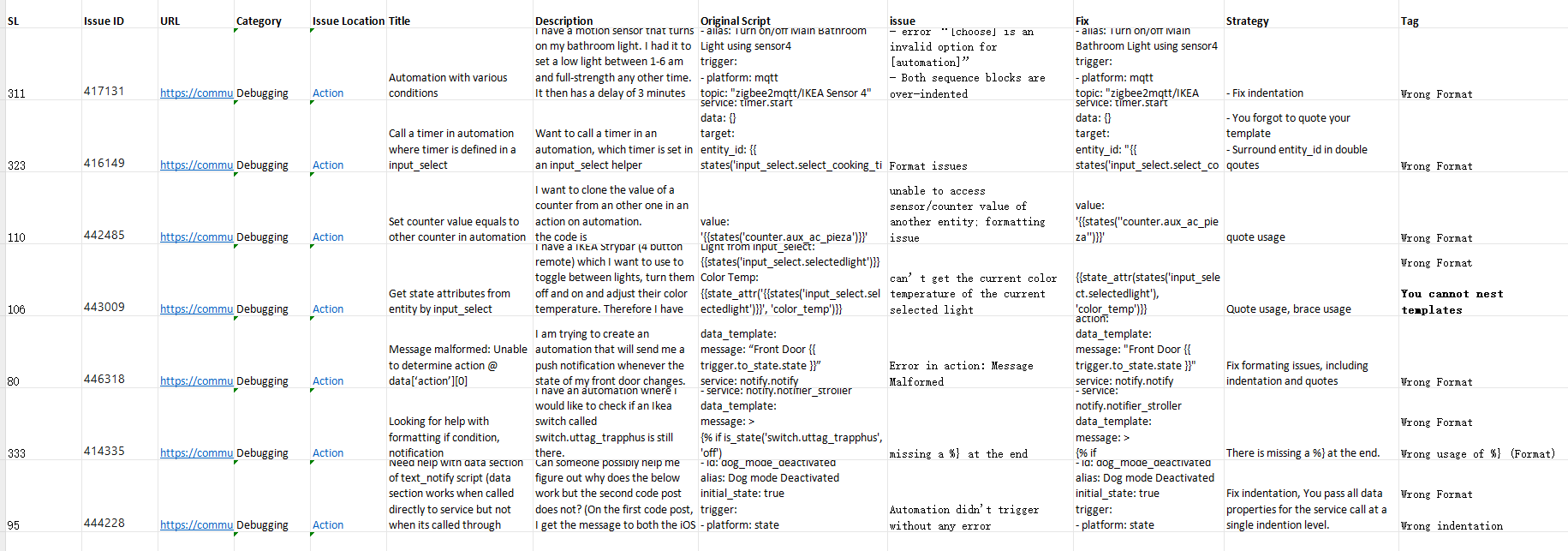}
    \caption{Overview of the debugging dataset.}
    \label{fig:debugging_data}
\end{figure*}

We extracted the dataset from anik at el’s work titled "Programming of Automation Configuration in Smart Home Systems: Challenges and Opportunities"\cite{Anik2018}. They crawled Home Assistant Community (HAC)\cite{web4}\cite{web3} and manually reviewed the discussion threads under the category “Configuration”. They considered threads tagged with “automation” from 03/01/2022 to 08/31/202222. Then they retrieved a dataset of 438 candidate discussion threads based on the HAC ranking and timestamp range. After a manual inspection of each thread, they finally selected 190 threads. Among them,129 (68\%) examined issues concerning debugging. Although existing tools can detect at most 14 issues and fix none. We collected their 129 debugging issues and did further investigation utilizing Large Language Models and different prompt engineering. The overview of the debugging dataset is presented in Figure~\ref{fig:debugging_data}.

The dataset comprises the debugging data related to automation configuration issues in smart home systems. As we manually evaluated different Large Language Models considering different prompt designs (Evaluated on 3 models generating 4 prompts), we randomly selected 21 data from them. 

\subsubsection{Representative Resolution Strategies}
\begin{figure*}[h!]
    \centering
    \includegraphics[width=0.88\textwidth]{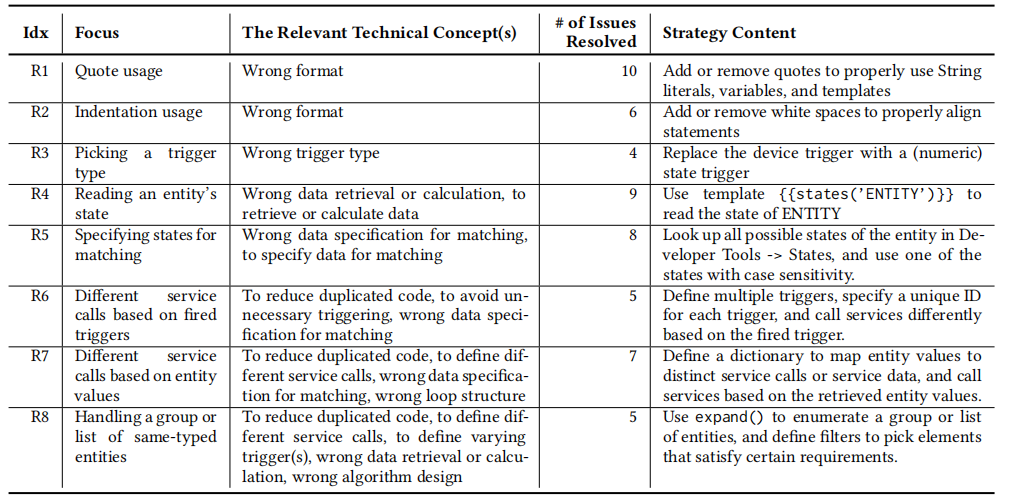}
    \caption{Representative Resolution Strategies.}
    \label{fig:eight_strategies}
\end{figure*}

Anik at el\cite{Anik2018} observed something insightful among the data they have crawled from HAC.  They identified some strategies (eight) that can be repetitively applied to resolve multiple issues of configuration bugs. They listed each strategy in terms of its focus, the relevant technical concepts, the number of issues it resolved, and a detailed description of the strategy's content shown in Figure \ref{fig:eight_strategies}. They also ranked the eight strategies in ascending order based on the number of relevant concepts. This structured information ensured clarity in understanding the purpose and applicability of each strategy, which we subsequently used as rules to map the strategies to the dataset. 
Our dataset is available here:  
\underline{\href{https://virginiatech-my.sharepoint.com/:x:/g/personal/msheikhmoonwaraa_vt_edu/Ea4C2y56wMBJkABfFdWnMuMBZYCPXryJwgWYFeHsZEOlOw?e=TlCn0X}{Configuration bug dataset with predicted fix and strategies}}.

\subsection{Prompt Engineering}
\subsubsection*{Prompt Design }
To evaluate how LLMs respond to different levels of contextual information and to test the models' ability to generalize across various configuration bug scenarios, we considered four types of prompts. These prompts were designed to guide LLMs in predicting strategies for resolving configuration bugs.

\begin{figure*}[h!]
    \centering
    \begin{subfigure}[b]{0.45\textwidth}
        \centering
        \includegraphics[width=\textwidth]{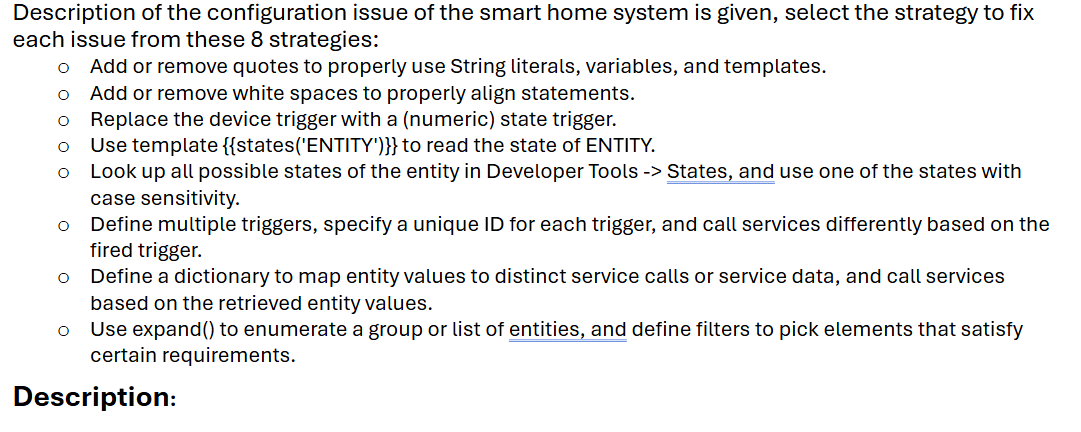}
        \caption{Prompt 1}
        \label{fig:sub1}
    \end{subfigure}
    \hfill
    \begin{subfigure}[b]{0.45\textwidth}
        \centering
        \includegraphics[width=\textwidth]{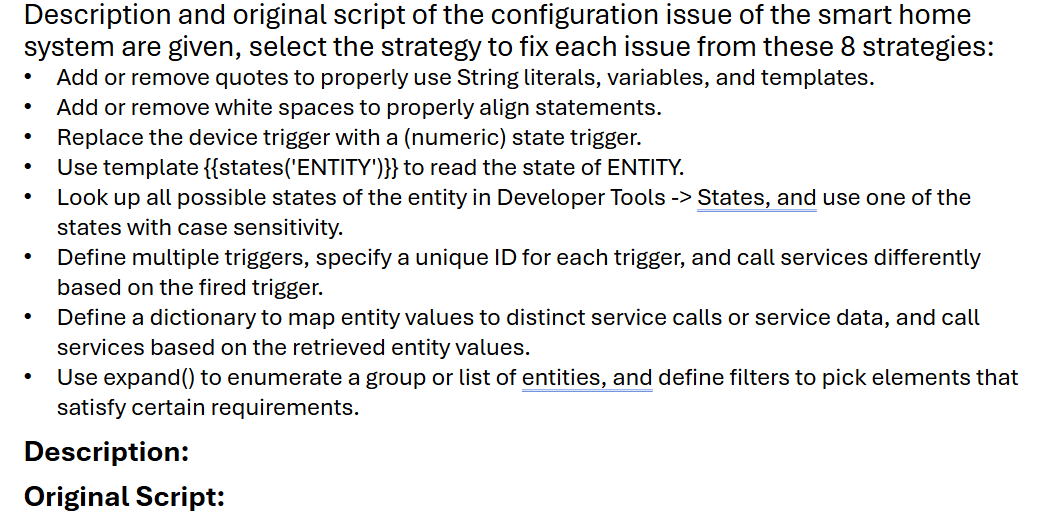}
        \caption{Prompt 2}
        \label{fig:sub2}
    \end{subfigure}

    \vskip 0.5cm 
    \begin{subfigure}[b]{0.45\textwidth}
        \centering
        \includegraphics[width=\textwidth]{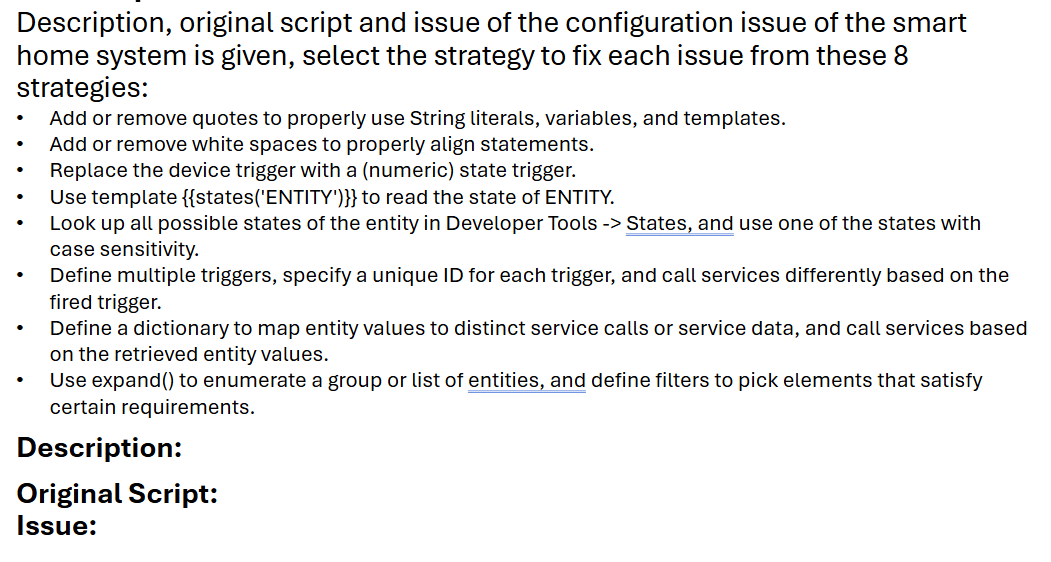}
        \caption{Prompt 3}
        \label{fig:sub3}
    \end{subfigure}
    \hfill
    \begin{subfigure}[b]{0.45\textwidth}
        \centering
        \includegraphics[width=\textwidth]{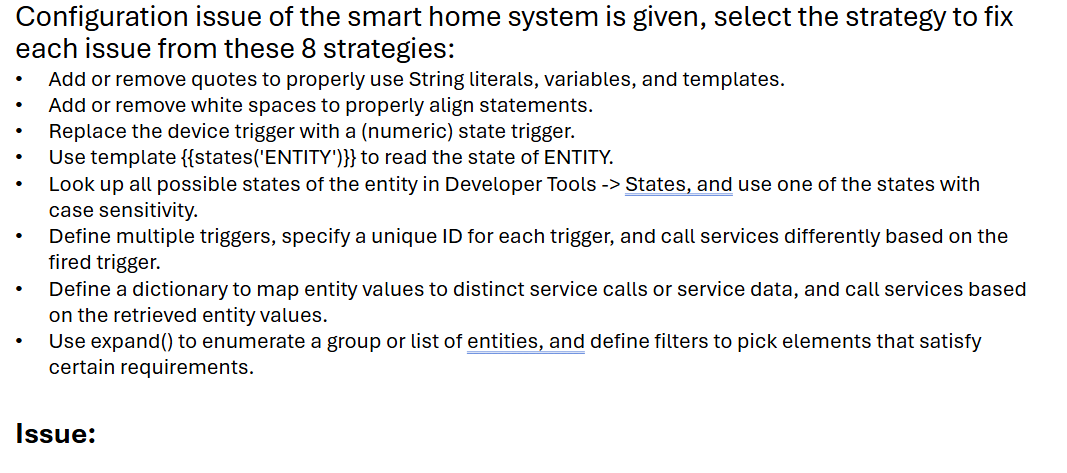}
        \caption{Prompt 4}
        \label{fig:sub4}
    \end{subfigure}

    \caption{Four prompt designs for generating fix strategies}
    \label{fig:four_prompts_grid}
\end{figure*}

We designed different prompts to provide information to the LLMs, enabling them to generate accurate predictions. The four distinct prompts are as follows:

\subsubsection{Prompt 1 (Description Only):}
This prompt provided only a description of the configuration issue without including additional details like the original script or specifics of the bug. However, the description of the configuration bug contained key information necessary to understand the issue. An example of Prompt 1 is shown in Figure \ref{fig:sub1}.

\subsubsection{Prompt 2 (Description + Original Script):}
This prompt offered more context to the model by combining the description with the original configuration script. It included both the description of the issue, which explained the problem in detail, and the script where the bug exists. This combination was intended to improve the model's accuracy in suggesting correct strategies and fixes. An example of Prompt 2 is shown in Figure \ref{fig:sub2}.

\subsubsection{Prompt 3 (Description + Original Script + Specific Issue):}
Prompt 3 included a detailed outline of the specific issue along with the description and the original script. In cases where the original script was too large or complex to understand, this prompt could be particularly effective in guiding the model to generate accurate predictions. An example of Prompt 2 is shown in Figure \ref{fig:sub3}.

\subsubsection{Prompt 4 (Issue Only):}
Prompt 4 focused solely on the specific issue without providing broader context or the original script. The purpose of this prompt was to evaluate how accurately the LLMs could predict strategies and fixes for resolving configuration bugs when only the specific issue was provided. An example of Prompt 2 is shown in Figure \ref{fig:sub4}.

\subsection{Selection of Large Language Models}
The application of Large Language Models (LLMs) in solving coding-related problems has been increasing significantly\cite{Naveed2023ACO}. In our approach, we considered three advanced and popular LLMs for our empirical evaluation, which were also accessible to us: GPT-4, GPT-4o, and Claude 3.5 Sonnet. These models were evaluated in this study to predict resolution strategies and fixes for configuration bugs in smart home systems.

These models possess advanced capabilities in understanding and generating code. The evaluation focused on mapping the predictions generated by the models to eight representative strategies and the suggested fixes. The Table\ref{tab:comparison_llm} highlights the differences between the selected LLMs.
\begin{table}[h!]
\centering
\caption{Comparison of GPT-4, GPT-4o, and Claude 3.5 Sonet}
\label{tab:comparison_llm}
\resizebox{\columnwidth}{!}{ 
\begin{tabular}{|l|c|c|c|}
\hline
\textbf{Feature}            & \textbf{GPT-4}       & \textbf{GPT-4o}       & \textbf{Claude 3.5}   \\ \hline
\textbf{Focus}              & Advanced reasoning   & Cost-effective, fast  & Safety, large contexts \\ \hline
\textbf{Applications}       & General-purpose      & Production-scale apps & Sensitive, aligned tasks \\ \hline
\textbf{Multimodal}         & Yes (limited)        & Yes (platform-based)  & Text only             \\ \hline
\end{tabular}}
\end{table}

\subsubsection{GPT4}
GPT-4 is a large language model developed by OpenAI, designed to excel in complex reasoning and natural language generation tasks. It performs exceptionally well in a variety of applications, including coding, content creation, and data analysis \cite{achiam2023gpt}.

The architecture of GPT4 builds upon GPT3.5, offering enhanced contextual understanding and fine-grained accuracy\cite{unknown}. These improvements make GPT4 particularly suitable for tasks that require in-depth problem-solving or nuanced responses\cite{sagodi2024reality}.

\subsubsection{GPT4o}
GPT4o is a cost-effective and optimized version of GPT4 developed by OpenAI\cite{10.1145/3650105.3652291}\cite{hurst2024gpt}. This model is also owned as GPT-4 Turbo. It is designed to deliver faster responses while maintaining high accuracy. This model is particularly ideal for scenarios that require scalability, such as handling high-volume queries, without significantly compromising performance.

Although GPT4 Turbo may lack some depth in addressing complex tasks compared to GPT4, it offers notable advantages in terms of speed, cost, and capability, making it a practical choice for many applications\cite{web1}.

\subsubsection{Claude 3.5 Sonnet}
 Claude 3.5 Sonnet: Created by Anthropic, this model emphasizes safety and the ability to handle large contexts, making it suitable for sensitive tasks\cite{web2}\cite{Enis2024FromLT}.

\subsection{Mapping to Representative Strategies} 
We utilized the representative strategies suggested by Anik et al.\cite{Anik2018}, which contain structured information to aid in understanding the purpose and applicability of each strategy, to map the strategies to the dataset. Based on these eight listed strategies, we relabeled the dataset strategies accordingly.

\subsection{Generating Configuration Fix Strategies and Fixes Using Prompts and LLMs}
Our approach utilized Large Language Models (LLMs) to generate strategies for fixing configuration bugs in smart home automation by leveraging our designed prompts. We also applied the same approach to generate fixes for configuration bugs utilizing Large Language Models (LLMs). These prompts were carefully crafted to guide the models in effectively identifying and resolving configuration issues.

\subsection{Compare different models and prompts}
After generating bug-fix strategies and the fixes for all 21 data points using four prompts and three large language models, we compared the outputs of the different models. Additionally, we analyzed the predictions of the models when using the same prompts.

\section{Implementation}
\subsection{Dataset Collection}
To conduct our work we considered the extracted 129 debugging issues and 8 representative resolution strategies from previous research on "Programming of Automation Configuration in Smart Home Systems: Challenges and Opportunities" \cite{Anik2018}. They collected the dataset from the Home Assistant Community(HAC)\cite{web4} and provided the eight representative resolution strategies with relevant information such focus of the strategy, the relevant technical concept(s), the number of issues it resolved, and a detailed description of the strategy's content.

\subsection{Reform Dataset}
We applied the eight representative resolution strategies, along with relevant information as rules, to map the strategies from the extracted dataset to these commonly used resolution strategies. The mapping for each issue was finalized through consensus between both authors.

\subsection{Generate prompts}
We generated four types of prompts based on the data available to us. These prompts were designed to include attributes that could provide insightful information for predicting the fix strategies and generating fixed scripts.

\subsection{Select Large Language Models}
We selected the large language models based on their advanced capabilities to read and generate code, and also their availability. We have considered three models, GPT4, GPT4o, and Claude 3.5 Sonnet, and implemented our approach using all four generated prompts.
Selecting the Large Language Models for predicting strategies to fix configuration bugs.

\subsection{Configuration Fix Strategies Generation}

\subsubsection{Input to LLMs}
We provided the models with the configuration bug details and the representative strategy list as input, using our predesigned prompts Figure~\ref{Example of prompt 2 as input to LLMs}.

\subsubsection{Output from LLMs}
The models analyzed the information provided through prompts. Then suggested fixing strategies from the provided strategy list as output.

\subsection{Configuration Fixes Generation}

\subsubsection{Input to LLMs}
\begin{figure}[h!]
    \centering
   
    \begin{subfigure}[b]{0.45\textwidth}
        \centering
        \includegraphics[width=\textwidth]{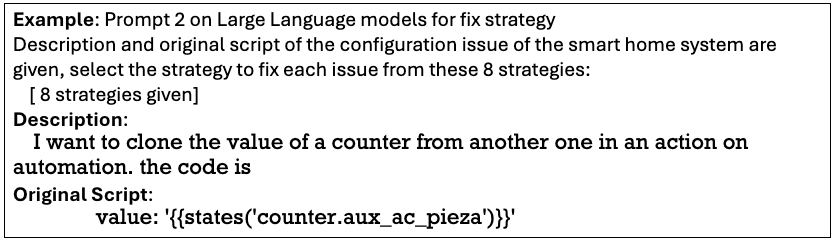}
        \caption{Example of prompt 2 as input to LLMs for fix strategies}
        \label{Example of prompt 2 as input to LLMs}
    \end{subfigure}

    \vskip 0.5cm 
    \begin{subfigure}[b]{0.45\textwidth}
        \centering
        \includegraphics[width=\textwidth]{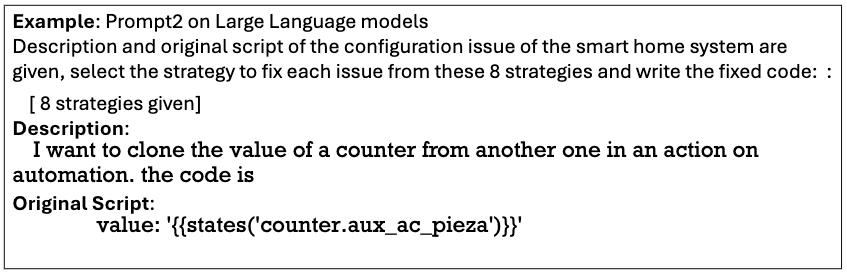}
        \caption{Example of prompt 2 as input to LLMs for fixes}
        \label{Example of prompt 2 fix as input to LLMs}
    \end{subfigure}

    \caption{Four prompt designs for generating fix strategies}
    \label{fig:four_prompts_grid}
\end{figure}

 Using our predesigned prompts, we provided the models with the configuration bug details as input. We specifically designed these prompts to provide the necessary context and guide the models generating fixed scripts figure~\ref{Example of prompt 2 fix as input to LLMs}.
 
\subsubsection{Output from LLMs}
The models analyzed the information provided through different prompts. They generated fixed scripts for the given configuration bugs and presented them as output along with reasoning.

\subsection{Compare Prompts}
After generating the fix strategies and solutions for configuration bugs using three models and all four prompts, we compared the performance of the prompts. We also analyzed the output from different models for the same prompt. While evaluating the suggested fixes, we focused only on cases where the models correctly predicted the strategies.

\subsection{Compare LLMs}
We evaluated the outcomes of the three models against the original strategies and fixes. Additionally, we compared the models' predictions for the same prompt.

The strategies of different prompts.

\section{Evaluation}
We completed all the implementation steps and collected the suggested strategies and fixes generated by different prompts for each configuration bug. Subsequently, we manually evaluated all the fix strategies and suggested fixes from GPT4, GPT4o, and Claude 3.5 Sonnet against the strategies derived from the dataset.

\subsection{High-level evaluation of the models}
\begin{figure*}[h!]
    \centering
    \includegraphics[width=0.98\textwidth]{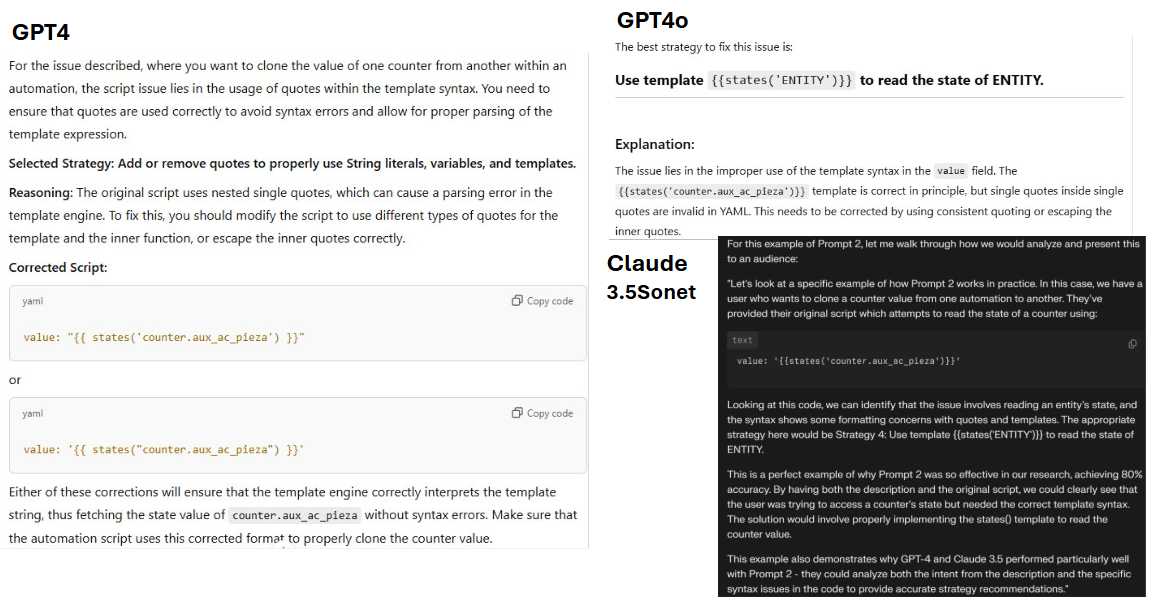}
    \caption{Example suggested fix strategies by three LLMs.}
    \label{fig:LLMsReasonings}
\end{figure*}

The applied large language models each have distinct advantages. GPT4 demonstrates superior reasoning capabilities compared to GPT4o. On the other hand, Claude 3.5 Sonnet offers better reasoning for suggesting strategies and fixes for configuration bugs than GPT4o, though it is not as comprehensive as GPT4. An example of suggested fix strategy by these three models Figure~\ref{fig:LLMsReasonings}

\subsection{Model Performance}
We evaluated all the large language models considered in this study based on the fix strategies and fixes they generated for configuration bugs in smart home systems.
\subsubsection{Predicting Fix Strategies }
Overall, GPT4 and Claude 3.5 Sonnet demonstrated the strongest performance with Prompt 2 (Description + Original Script), achieving 80\% accuracy in strategy prediction for fixing smart home system configuration bugs. GPT-4, however, exhibited more consistent results across different prompt types.
The performance of GPT-4 was strong with Prompt 2 and Prompt 3 but poor with Prompt 1 and Prompt 4. Claude 3.5 Sonnet performed particularly well with Prompt 3. All three models demonstrated similar performance when using Prompt 4. Tables \ref{tab:prompt1}, \ref{tab:prompt2}, \ref{tab:prompt3}, and \ref{tab:prompt4} show the performance of the three models utilizing the four selected prompts.

GPT-4 and GPT-4o provided 52\% similar strategies when using Prompts 1 and 2. The similarity dropped to 38\% for Prompts 3 and 4, as shown in Table \ref{tab:similar_suggestions}.

\begin{table}[h!]
\centering
\caption{Performance of LLMs predicting Fixes Strategies using Prompt 1}
\label{tab:prompt1}
\begin{tabularx}{\columnwidth}{|X|X|X|X|X|X|X|}
\hline
      Total       & GPT4 & GPT4o & Claude &  GPT4 \% & GPT4o \% &  Claude \% \\ \hline
21        & 8    & 11    & 8      & 38\%    & 52\%     & 38\%      \\ \hline
\end{tabularx}
\end{table}

\begin{table}[h!]
\centering
\caption{Performance of LLMs predicting Fixes Strategies using Prompt 3}
\label{tab:prompt2}
\begin{tabularx}{\columnwidth}{|X|X|X|X|X|X|X|}
\hline
      Total    & GPT4 & GPT4o & Claude & GPT4 \% & GPT4o \% & Claude \% \\ \hline
     21   & 17   & 10    & 11     & 80\%    & 47\%     & 52\%      \\ \hline
\end{tabularx}
\end{table}

\begin{table}[h!]
\centering
\caption{Performance of LLMs predicting Fixes Strategies using Prompt 3}
\label{tab:prompt3}
\begin{tabularx}{\columnwidth}{|X|X|X|X|X|X|X|}
\hline
      Total       & GPT4 & GPT4o & Claude & GPT4 \% & GPT4o \% &  Claude \% \\ \hline
     21   & 15   & 14    & 17     & 71\%    & 67\%     & 80\%      \\ \hline
\end{tabularx}
\end{table}

\begin{table}[h!]
\centering
\caption{Performance of LLMs predicting Fixes Strategies using Prompt 3}
\label{tab:prompt4}
\begin{tabularx}{\columnwidth}{|X|X|X|X|X|X|X|}
\hline
      Total       & GPT4 & GPT4o & Claude & GPT4 \% & GPT4o \% & Claude \% \\ \hline
    21    & 9    & 9     & 10     & 42\%    & 42\%     & 47\%      \\ \hline
\end{tabularx}
\end{table}

\begin{table}[h!]
\centering
\caption{Similar Fix Strategies from GPT4 \& GPT4o}
\label{tab:similar_suggestions}
\renewcommand{\arraystretch}{1.6} 
\setlength{\tabcolsep}{1pt}      
\resizebox{\columnwidth}{!}{     
\begin{tabular}{|l|c|c|c|c|}
\hline
\textbf{Prompts}     & \textbf{Prompt1} & \textbf{Prompt2} & \textbf{Prompt3} & \textbf{Prompt4}  \\ \hline
\textbf{Similar suggestion from \textbf{GPT4} \& \textbf{GPT4o}} & 11               & 11               & 8                & 8                \\ \hline
\textbf{Percentage}            & 52\%             & 52\%             & 38\%             & 38\%             \\ \hline
\end{tabular}}
\end{table}

\subsubsection{Predicting Fixes}

By observing the predicted strategies for fixing configuration bugs across different models and prompts, we measured the number of matched strategies and the proportion of correct fixes generated by the models. During the evaluation, some fixes could not be definitively assessed for correctness and were labeled as Do Not Know.

From the tables\ref{tab:performance_summary}, \ref{tab:performance_p2}, \ref{tab:performance_p3},\ref{tab:performance_p4} it can be concluded that the overall performance of GPT4 is better than the other two models. While GPT4o performed poorly with Prompt 1, it showed good performance with the other three prompts. Claude 3.5 Sonnet performed well with Prompts 3 and 4; however, it provided some incorrect fixes in all cases.

\begin{table}[h!]
\centering
\caption{Performance on predicting Fixes using Prompt 1 }
\label{tab:performance_summary}
\renewcommand{\arraystretch}{1.3} 
\setlength{\tabcolsep}{8pt}      
\resizebox{\columnwidth}{!}{     
\begin{tabular}{|l|c|c|c|}
\hline
LLMs & \textbf{GPT4} & \textbf{GPT4o} & \textbf{Claude} \\ \hline
Matched Strategies             & 7                 & 11 & 9               \\ \hline
Correct              & 6                 & 6  & 5              \\ \hline
Percentage        & 85.71\%           & 54.55\%  & 55.56\%          \\ \hline
Do Not Know       & 1                 & 2  & 2                \\ \hline
Incorrect         & 0                 & 3  & 3            \\ \hline
Percentage        &0 \%           & 27.2727\%   & 27.27\%        \\ \hline
\end{tabular}}
\end{table}

\begin{table}[h!]
\centering
\caption{Performance on predicting Fixes using Prompt 2}
\label{tab:performance_p2}
\resizebox{\columnwidth}{!}{ 
\begin{tabular}{|l|c|c|c|}
\hline
LLMs & \textbf{GPT4} & \textbf{GPT4o} & \textbf{Claude}\\ \hline
Matched Strategies    & 17                & 10 &  9            \\ \hline
Correct      & 14                & 9 & 5               \\ \hline
Percentage  & 82.35\%           & 90\% & 55.56\%             \\ \hline
Do Not Know & 1                 & 0     & 0          \\ \hline
Incorrect  & 2                 & 1        &4     \\ \hline
Failure \% & 11.76\%           & 10\%      &44.44\%        \\ \hline
\end{tabular}}
\end{table}

\begin{table}[h!]
\centering
\caption{Performance on predicting Fixes using Prompt 3}
\label{tab:performance_p3}
\resizebox{\columnwidth}{!}{ 
\begin{tabular}{|l|c|c|c|}
\hline
LLMs   & \textbf{GPT4} & \textbf{GPT4o}  & \textbf{Claude} \\ \hline
Matched Strategies     & 15                & 14  & 15            \\ \hline
Correct       & 14                & 12   & 13           \\ \hline
Percentage & 93.33\%           & 85.71\% & 86.67\%         \\ \hline
Do Not Know & 1                 & 2    & 1           \\ \hline
Incorrect  & 0                 & 0    & 1           \\ \hline
Failure \% & 0\%               & 0\%  & 6.67\%            \\ \hline
\end{tabular}}
\end{table}

\begin{table}[h!]
\centering
\caption{Performance on predicting Fixes using Prompt 4}
\label{tab:performance_p4}
\resizebox{\columnwidth}{!}{ 
\begin{tabular}{|l|c|c|c|}
\hline
LLMs   & \textbf{GPT4} & \textbf{GPT4o} & \textbf{Claude} \\ \hline
Matched Strategies      & 9      & 9     & 9            \\ \hline
Correct       & 7               & 7      & 7          \\ \hline
Percentage & 77.78\%          & 77.78\%   & 77.78\%        \\ \hline
Do Not Know & 2               & 0         & 0     \\ \hline
Incorrect  & 0                 & 2        & 2      \\ \hline
Failure \% & 0\%           & 22.22\%      & 22.22\%     \\ \hline
\end{tabular}}
\end{table}

\subsection{Prompt Evaluation }

\subsubsection{Predicting Fix Strategies }
We evaluated the different prompts based on the number of fix strategies the models predicted using them. Overall, Prompt 3 and Prompt 4 provided better suggestions than the other two prompts for selecting strategies to fix configuration bugs. 

Table 3 shows that all three models performed well with Prompt 3, where each model provided the correct strategy for more than 67\% of the total data. Among them, Claude 3.5 Sonnet outperformed GPT4 and GPT4o, even though it did not perform as well with the other prompts. 

GPT4 performed well with Prompt 2 but provided poor suggestions with Prompt 1 and Prompt 4.
Prompt 4 did not perform well with any of the three models. None of the models were able to correctly predict even half of the strategies for fixing bugs.

\subsubsection{Predicting Fixes}
We measured the number of matched strategies and how many of these included correct fixes generated by the models. During the evaluation of the fixes predicted by the models, there were instances where we could not determine their correctness, and we labeled these cases as Do Not Know.

For Prompt 2, GPT4 correctly predicted 85.71\% of the fixes from the correct strategies it provided, with only one fix labeled as Do Not Know. On the other hand, GPT-4o produced more than 27\% incorrect fixes among the correct strategies it provided. Table \ref{tab:performance_summary} presents a detailed comparison.

GPT-4o outperformed the other models when using Prompt 2. Table \ref{tab:performance_p2} shows that GPT-4o correctly predicted 90\% of the fixes from the correct strategies it provided, with only one fix labeled as Do Not Know. However, GPT-4 and GPT-4o produced 11.76\% and 10\% incorrect fixes, respectively. Table \ref{tab:performance_p2} presents the details.

For Prompt 3, GPT-4 excelled in providing correct fixes for configuration bugs, achieving more than 93\% accuracy among the fix strategies it predicted correctly. GPT-4o also performed well, delivering approximately 86\% correct fixes, although some of its other suggested fixes were labeled as Do Not Know. Table \ref{tab:performance_p3} presents a detailed comparison.

The table \ref{tab:performance_p4} shows that GPT4 and GPT4o performed equally well in predicting fixes for configuration bugs correctly. Both models provided approximately 78\% of the fixes accurately from the correct strategies they predicted. However, GPT-4o also suggested 22.22\% of the fixes incorrectly.

\section{Related work}
Several research efforts have explored ways to improve configuration challenges processes through automation and tool support.

The work of Duc et al. \cite{10.1145/3120459.3120471} examined the main failure patterns of smart home systems, this work focuses on hardware failure(wireless link loss, battery damage, power outage). Chen et al. \cite{Chen2017ApplicationOF} analyzed the fault symptoms and provided maintenance suggestions after modeling the IoT system with four layers: application, storage, communication, and data.

In the realm of natural language processing for code, \cite{10.1145/3212695} surveyed various techniques for learning from source code, including approaches for code summarization and bug detection. Their work provides a foundation for our use of NLP and machine learning to analyze code changes and generate human-readable comments. \cite{article}

V. J. HELLENDOORN ET AL\cite{hellendoorn2020global} explored the use of machine learning models(RNN, Transformer, GGRN etc.)\cite{bani2021deep,han2021transformer,tkachenko2020approach} for code completion and bug detection. Their findings on the effectiveness of transformer-based models for code understanding tasks inform our choice of model architectures for comment generation. However, they did not consider any large language model for analysis.

A few recent studies have focused on identifying and characterizing issues in IoT systems \cite{8835392}\cite{10.1145/3290605.3300782}. Makhshari and Mesbah \cite{9402092} conducted interviews and surveys with IoT developers, revealing that testing and debugging are the primary challenges faced. However, their work does not address configuration issues. Brackenbury et al. \cite{10.1145/3290605.3300782} focused on the trigger-action programming (TAP) model. They analyzed and systematized temporal paradigms within TAP systems. Their work identified that TAP systems express rules and categorized TAP programming bugs into three main types: control logic errors, timing issues, and bugs arising from inaccurate user expectations. Ahmad et al. \cite{10462177} designed and implemented a framework to quantitatively evaluate how effectively an LLM can fix specified bugs related to hardware security.

Focusing on root causes, fixes, triggers, and impacts Wang et al. \cite{10.1145/3533767.3534365} analyzed 330 device integration bugs from HAC. This work is similar to our work, rather than addressing device integration, our study concentrates on coding issues in automation configuration.

S. M. H. Anik et al. \cite{Anik2018} conducted a comprehensive study on the challenges and opportunities in programming automation configurations for smart home systems \cite{inproceedings}. Their work provides valuable insights into the complexities of setting up and maintaining smart home devices, which has informed our approach to developing AI-assisted configuration tools. Our study was conducted using the dataset extracted in their work.

In the realm of IoT security, the Bitdefender 2024 IoT Security Landscape Report\cite{Bitdefender} offers crucial insights into the evolving threats facing smart homes. This report highlights the need for robust security measures in smart home configurations, which our AI-assisted approach aims to address.
 
Although none of these previous works focused on the evaluation of LLMs in predicting
fixes of Configuration bugs in Smart Home System, these insights help us understand the problem fully and design our solution design considering it.

\section{Conclusion}
GPT-4 and Claude 3.5 Sonnet demonstrated exceptional capabilities in predicting fix strategies for smart home configuration bugs, achieving 80\% accuracy when provided with both bug descriptions and original scripts[1]. GPT-4 maintained consistent performance across different prompt types, while GPT-4o offered advantages in processing speed and cost-effectiveness despite slightly lower accuracy rates.

The research revealed that prompt design significantly influences model performance. Prompt 2 (Description + Original Script) and Prompt 3 (Description + Original Script + Specific Issue) consistently outperformed other designs across all models[1]. This finding suggests that providing comprehensive context is crucial for accurate bug fix prediction. GPT4 achieved remarkable accuracy in fix generation, reaching 93.33\% success rate with Prompt 3, while GPT-4o achieved 90\% accuracy with Prompt 2[1]. These results demonstrate the models' strong capability in not only identifying appropriate fix strategies but also generating correct solutions. This study introduces a novel framework for evaluating LLMs in the context of smart home configuration debugging. The development of four distinct prompt types and their systematic evaluation provides valuable insights for future research in automated bug fixing. The findings offer immediate practical value for smart home system developers and users. The high accuracy rates achieved by the models suggest that LLM-based tools could significantly streamline the debugging process in smart home configurations. Future research should focus on improving model performance for complex configuration scenarios and reducing the occurrence of incorrect fix suggestions. Particular attention should be paid to cases where models currently produce uncertain or incorrect results. The promising results suggest opportunities for developing automated debugging tools that integrate LLMs with existing smart home platforms. Such integration could provide real-time assistance to users encountering configuration issues. Further investigation into prompt design optimization could yield even better results. Special attention should be given to reducing the 22.22\% failure rate observed in some scenarios with GPT4o. While the study demonstrates significant progress in automated bug fixing for smart home systems, some limitations remain. The research was conducted on a dataset of 21 randomly selected cases from a larger pool of 129 debugging issues[1]. Future work with larger datasets could provide more comprehensive insights into model performance across a broader range of configuration scenarios.


\section*{Acknowledgment}
We are grateful to our professor, Na Meng, in the Department of Computer Science at Virginia Tech for their insightful feedback and support throughout the research process. The authors would like to express their sincere gratitude to Anik et al. for providing the valuable dataset used in this study. Finally, we acknowledge the anonymous reviewers for their constructive comments that helped improve the quality of this paper.



%
\bibliographystyle{siam}
\bibliography{references}

\end{document}